\documentclass[prl,showpacs,superscriptaddress,twocolumn]{revtex4}

\usepackage{amsfonts}
\usepackage{amsmath}
\usepackage{amssymb}
\usepackage{graphicx}

\begin{document}

\title{Exact time-reversal focusing of acoustic and quantum excitations in open cavities: The perfect inverse filter}

\author{Hern\'an~L. Calvo}
\affiliation{Instituto de F\'{\i}sica Enrique Gaviola (CONICET) and Facultad de Matem\'{a}tica, Astronom\'{\i}a y F\'{\i}sica, Universidad Nacional de C\'{o}rdoba, Ciudad Universitaria, 5000 C\'{o}rdoba, Argentina.}
\author{Horacio~M.~Pastawski}
\affiliation{Instituto de F\'{\i}sica Enrique Gaviola (CONICET) and Facultad de Matem\'{a}tica, Astronom\'{\i}a y F\'{\i}sica, Universidad Nacional de C\'{o}rdoba, Ciudad Universitaria, 5000 C\'{o}rdoba, Argentina.}

\begin{abstract}
The time-reversal mirror (TRM) prescribes the reverse playback 
of a signal to focalize an acoustic excitation as a Loschmidt echo. In 
the quantum domain, the perfect inverse filter (PIF) processes this signal 
to ensure an exact reversion provided that the excitation originated outside 
the cavity delimited by the transducers. We show that PIF takes a simple 
form when the initial excitation is created inside this cavity. This also 
applies to the acoustical case, where it corrects the TRM and improves the 
design of an acoustic bazooka. We solve an open chaotic cavity modeling a 
quantum bazooka and a simple model for a Helmholtz resonator, showing that 
the PIF becomes decisive to compensate the group velocities involved in a 
highly localized excitation and to achieve subwavelength resolution.
\end{abstract}


\maketitle

\section{Introduction}

The time-reversal mirror (TRM) technique was developed by Mathias Fink and
collab. \cite{FinkSA} in order to refocus an ultrasonic excitation
diffusing away through a heterogeneous medium. The remarkable robustness of
this technique against inhomogeneities and imperfections in the procedure
enabled novel applications in medical physics \cite{Montaldo}, oceanography 
\cite{Edelman} and telecommunications \cite{Henty}. The acoustic signals
emerging from a localized source point are recorded by an array of
transducers. Ideally, this array splits the system in two portions: the
internal cavity, where the focusing takes place, and the outer region,
towards where the excitation finally spreads away. The transducers then
play-back each registry in the inverted time sequence. Surprisingly, this
simple prescription allows recovering the initial excitation with high
precision. This constitutes an alternative form of achieving a Loschmidt
Echo \cite{Jalabert}, \textit{i.e.} the time reversal of an excitation
distributed in a complex medium. To ensure the exact reversed
dynamics, we exploited the close correspondence between the Schr\"{o}dinger
and the classical wave equations \cite{DeRosny, CalvoPB} by introducing a
formal prescription denoted as the perfect inverse filter (PIF) \cite%
{PastawskiEPL}. This yields the precise injection function that produces the
targeted dynamics inside the cavity. In particular, the PIF prescribes how
to process the recorded signal in order to achieve the exact refocusing of
whatever excitation exists in the cavity, provided that it was originated as
a wave packet incoming from the outer region. A potential application is the 
\textit{quantum bazooka}, analogous to the \textit{acoustic bazooka}
conceived by M. Fink and collaborators \cite{Kuperman}. When a waveguide
injects a well-defined wave packet into a chaotic cavity, it is reflected as
a persistent noisy signal. The time-reversed counterpart allows to store in
the cavity the weak signal injected through a long period of time until it
emerges as a burst of excitation.

To further optimize the spatial contrast of a localized acoustic excitation
in cases where TRM results imperfect, Fink and collab. proposed the
spatio-temporal inverse filter (STIF) \cite{Tanter}. It requires the
inversion of a matrix built from the Fourier transforms of the direct
wavefield propagators. These give the response at the transducers to a Dirac
delta signal in each control point within the focal region. Unlike the TRM,
the STIF results in a highly invasive technique. Minimally invasive STIF
(miSTIF), a recent proposal by \textit{Vignon et al.} \cite{Vignon},
achieves the same focusing quality as STIF without resorting to control
points. All the information in the wavefield propagator is deduced from the
backscattering signals of the transducers and the technique results no more
invasive than TRM. Both the miSTIF and the PIF involve the inversion of a
propagator matrix in the reduced basis of the transducers. However, this may 
result in numerical instabilities that require a delicate regularization 
procedure to deal with the Green's function divergencies.

In this work, we assume a natural separation between the incoming and
outgoing excitation components, as when the source is internal, to develop a
PIF procedure based on the detection of the escaping wave. This internal
PIF results very simple and intrinsically stable. Moreover, it reveals a
property not obvious before: the filter that corrects the TRM does not
depend on the internal scattering, but only on the group velocities of the
various propagating modes in the far field region. This is general enough to
cover, either exactly or as an approximation, a wide range of TRM set-ups
that would benefit from a PIF. We may mention subwavelength focusing \cite%
{FinkSCIENCE}, lithotripsy in dispersive media \cite{Montaldo}, the
quantification of nonlinear media \cite{Ulrich}, coherent control of optical
excitation energies in nanoplasmonic systems \cite{Stockman} and acoustic
bazookas \cite{Kuperman}. To this list we may add the development of novel
coherent control strategies that seek the local injection of designed
quantum excitations. As discussed below, this involves fields as diverse as
nanoelectromechanics, Bose-Einstein condensates and NMR.

We start by considering a quantum system where the corresponding propagators
are well known \cite{CalvoPRL}. Then, we will show that the resulting PIF
prescription remains valid to describe sound propagation. A numerical
contrast with the TRM procedure in two situations of physical interest, the
quantum bazooka configuration and a simplified model for a Helmholtz
resonator coupled to a grated waveguide, confirms the relevance of the PIF
correction.

\section{Time-reversal in a cavity}

We consider the situation in which, at the initial time, the excitation is
concentrated inside the cavity. From $t_{0}=0$ until a registration time $%
t_{R}$, the discrete wave function%
\begin{equation}
\psi _{s}(t)=\sum\nolimits_{n}\mathrm{i}\hbar G_{s,n}^{R}(t)\psi _{n}(0),%
\text{ }0\leq t\leq t_{R},  \label{eq_for}
\end{equation}%
is detected and recorded as it escapes through the frontier. In the present
letter we consider a cavity, either integrable or chaotic, open through a 1d
channel. Therefore, its frontier is controlled by a single transducer at $%
x_{s}=sa$, with $s$ integer and $a$ the distance between neighbors sites in
a lattice. Although apparently a restriction, this simple model faithfully
represents the essential underlying physics. A straightforward
generalization requires a matrix formulation and will be presented elsewhere.

Here, $G_{m,n}^{R}(t)$ is the retarded Green's function relating the
amplitude response at $x_{m}$ when a Dirac delta signal $\delta (t)$ is
injected at $x_{n}$ inside the cavity. It satisfies the inhomogeneous Schr%
\"{o}dinger equation%
\begin{equation}
\mathrm{i}\hbar \frac{\partial }{\partial t}G_{m,n}^{R}(t)-\sum%
\nolimits_{i}H_{m,i}G_{i,n}^{R}(t)=\delta _{m,n}\delta (t),  \label{eq_green}
\end{equation}%
with $H_{m,i}$ the Hamiltonian matrix elements. Assuming that there are no
localized states, the registration time $t_{R}$ is taken long enough in
order to ensure that all excitations have left the cavity.

As we have shown in ref.\cite{PastawskiEPL}, the time reversion for a wave
packet arriving at the cavity from the outer region is performed, in the
energy domain, through the injection%
\begin{equation}
\chi _{s}(\varepsilon )=\frac{1}{\mathrm{i}\hbar G_{s,s}^{R}(\varepsilon )}%
\psi _{s}^{\mathrm{rev}}(\varepsilon ).
\end{equation}%
Notice that in the above expression,%
\begin{equation}
\psi _{s}^{\mathrm{rev}}(\varepsilon )\equiv \int_{-\infty }^{\infty }%
\mathrm{d}te^{\mathrm{i}\varepsilon t/\hbar }\psi _{s}^{\mathrm{rev}}(t),
\end{equation}%
accounts for the time reversed complete evolution of the wave packet, 
\textit{i.e.} both the incoming and outgoing components are required in
order to compute $\psi _{s}^{\mathrm{rev}}(t)$. In the present work we want
to reverse the signal that is produced inside the cavity. Then, we deal with
the building up of the complete evolution from the knowledge of $\psi
_{s}(t) $ at times $0\leq t\leq t_{R}$. On the assumption that the injection
function is known, the reversed propagation at the transducer would be%
\begin{align}
\psi _{s}^{\mathrm{rev}}(t)& =\psi _{s}^{\ast }(2t_{R}-t) \\
& =\mathrm{i}\hbar \int_{0}^{t}G_{s,s}^{R}(t-t^{\prime })\chi
(x_{s},t^{\prime })\mathrm{d}t^{\prime },
\end{align}%
for times $t_{R}\leq t\leq 2t_{R}$ before the focalization. Since the PIF
prescription achieves exact reversion inside the cavity, the evolution at
subsequent times ($t>2t_{R}$) can be conceived as the wave packet that
starts at the focalization time with $\psi _{n}(2t_{R})=\psi _{n}^{\ast }(0)$%
. Thus, for subsequent times we expect%
\begin{equation}
\psi _{s}^{\mathrm{rev}}(t)=\sum\limits_{n}\mathrm{i}\hbar
G_{s,n}^{R}(t-2t_{R})\psi _{n}^{\ast }(0).  \label{eq_subs}
\end{equation}%
A comparison between eqs.(\ref{eq_for}) and (\ref{eq_subs}) shows that for
the most simple case in which all coefficients $\psi _{n}(0)$ are real, both
evolutions become identical. The same evolution but sign changed should be
obtained when they are purely imaginary. In a more general situation where
the coefficients present different phases we start with the Fourier transform%
\begin{align}
\psi _{s}^{\mathrm{rev}}(\varepsilon )& =\int_{t_{R}}^{2t_{R}}\psi
_{s}^{\ast }(2t_{R}-t)e^{\mathrm{i}\varepsilon t/\hbar }\mathrm{d}t+ \\
& +\mathrm{i}\hbar \sum\limits_{n}\int_{2t_{R}}^{\infty
}G_{s,n}^{R}(t-2t_{R})\psi _{n}^{\ast }(0)e^{\mathrm{i}\varepsilon t/\hbar }%
\mathrm{d}t \\
& =e^{\mathrm{i}\varepsilon 2t_{R}/\hbar }\left[ \psi _{s}^{\ast
}(\varepsilon )+\mathrm{i}\hbar \sum\limits_{n}G_{s,n}^{R}(\varepsilon )\psi
_{n}^{\ast }(0)\right] ,  \label{eq_rev}
\end{align}%
where the second term within the brackets can be interpreted as the unknown
evolution for subsequent times. The $\varepsilon $ argument and the
superscript in $G_{s,n}^{R}(\varepsilon )$ are omitted for a compact
notation henceforth. At this point, we account for the hopping terms $%
V_{s,s-1}$ and $V_{s-1,s}$ (with $x_{s-1}$ inside de cavity) connecting the
cavity with the outer region. Hence, we use the Dyson equation \cite%
{Economou} to rewrite%
\begin{equation}
G_{s,n}=G_{s,n}^{(0)}+G_{s,s}V_{s,s-1}G_{s-1,n}^{(0)},
\end{equation}%
where $G^{(0)}$ follows eq.(\ref{eq_green}) for the unperturbed Hamiltonian
with $V_{s,s-1}=V_{s-1,s}=0$. Since $s$ and $n$ are at disconnected
subsystems $G_{s,n}^{(0)}=0$. Besides, since the cavity subsystem is closed
and non absorbing, $V_{s,s-1}G_{s-1,n}^{(0)}$ is a real number. Therefore, we
obtain for the unknown evolution 
\begin{subequations}
\begin{equation}
\mathrm{i}\hbar \sum\limits_{n}G_{s,n}\psi _{n}^{\ast }(0)=-\frac{G_{s,s}}{%
G_{s,s}^{\ast }}\psi _{s}^{\ast }(\varepsilon ).
\end{equation}%
Finally, we use this last expression to rewrite eq.(\ref{eq_rev}) as 
\end{subequations}
\begin{equation}
\psi _{s}^{\mathrm{rev}}(\varepsilon )=e^{\mathrm{i}\varepsilon 2t_{R}/\hbar
}\left[ 1-\frac{G_{s,s}}{G_{s,s}^{\ast }}\right] \psi _{s}^{\ast
}(\varepsilon ).
\end{equation}%
Since we have the complete $\psi _{s}^{\mathrm{rev}}(\varepsilon )$, the
injection function according to the PIF formula is given by%
\begin{align}
\chi _{s}(\varepsilon )& =\frac{1}{\mathrm{i}\hbar G_{s,s}^{R}}\psi _{s}^{%
\mathrm{rev}}(\varepsilon ) \\
& =\frac{2}{\hbar }e^{\mathrm{i}\varepsilon 2t_{R}/\hbar }\mathrm{Im}\left(
G_{s,s}^{-1}\right) \psi _{s}^{\ast }(\varepsilon ).  \label{eq_PIF}
\end{align}%
This is the PIF prescription for the case where the initial state is an
excitation inside the cavity. As we shall see, the imaginary component of
the inverse in $G_{s,s}$ is closely related with the group velocity of the
scattered waves in the outer region. Therefore, contrary to the external
source condition presented in ref. \cite{PastawskiEPL}, the correction
imposed by the PIF in internal source condition does not depend on the
structure of the cavity.

In this Letter, we used the one-dimensional picture for simplicity.
Following the same reasoning as before, a multichannel generalization for
the PIF procedure implies minor details and will be described elsewhere. In
such case, one should be able to compute the relevant Green's function
components linking the transducers in the boundary through a matrix
continued fraction algorithm \cite{PastawskiPRB}. An alternative, more
general formulation for the PIF may be obtained based on the fundamental
optical theorem \cite{FoaTorres}. In words, it states that the continuous
density of states at a given site $x_{0}$ is built upon the escape rates
through the boundaries of the system. If we choose a semi-infinite chain and
there is no escape to the left of site $x_{0}$, the excitation can only
escape to the right of the frontier $x_{s}$ with a rate $\mathrm{Im}\Sigma
_{s}/\hbar $. In equations:%
\begin{equation}
\mathrm{Im}G_{0,0}=G_{0,s}\mathrm{Im}\Sigma _{s}G_{s,0}^{\ast }.
\end{equation}%
Multiplying both sides by a pulse $\xi (\varepsilon )$ and the $\mathrm{i}%
\hbar $ factor we find%
\begin{equation}
\mathrm{i}\hbar \left( G_{0,0}-G_{0,0}^{\ast }\right) \xi (\varepsilon )=2%
\mathrm{i}G_{0,s}\Gamma _{s}(-\mathrm{i}\hbar )G_{s,0}^{\ast }\xi
(\varepsilon ),  \label{eq_optic}
\end{equation}%
with $\Gamma _{s}=-\mathrm{Im}\Sigma _{s}$. Since the outgoing wave
originated at the focal point writes%
\begin{equation}
\psi _{0}(\varepsilon )=\mathrm{i}\hbar G_{0,0}\xi (\varepsilon ),
\end{equation}%
we re-obtain the PIF formula by identifying the second term in the left hand
of eq.(\ref{eq_optic}) as the incoming wave that produced the pulse in the
focal point. The complete evolution, identical to the reversed one, results%
\begin{align}
\psi _{0}(\varepsilon )+\psi _{0}^{\ast }(\varepsilon )& =\psi _{0}^{\mathrm{%
rev}}(\varepsilon ) \\
& =\mathrm{i}\hbar G_{0,s}\left[ \frac{2}{\hbar }\Gamma _{s}\psi _{s}^{\ast
}(\varepsilon )\right] .
\end{align}%
The term between the square brackets is the excitation that must be injected
at $x_{s}$ to reproduce the original signal in the cavity and coincides with
the PIF prescription obtained in eq.(\ref{eq_PIF}).

\subsection{PIF in a quantum bazooka configuration}

\begin{figure}[tbp]
\includegraphics[angle=90,width=8cm]{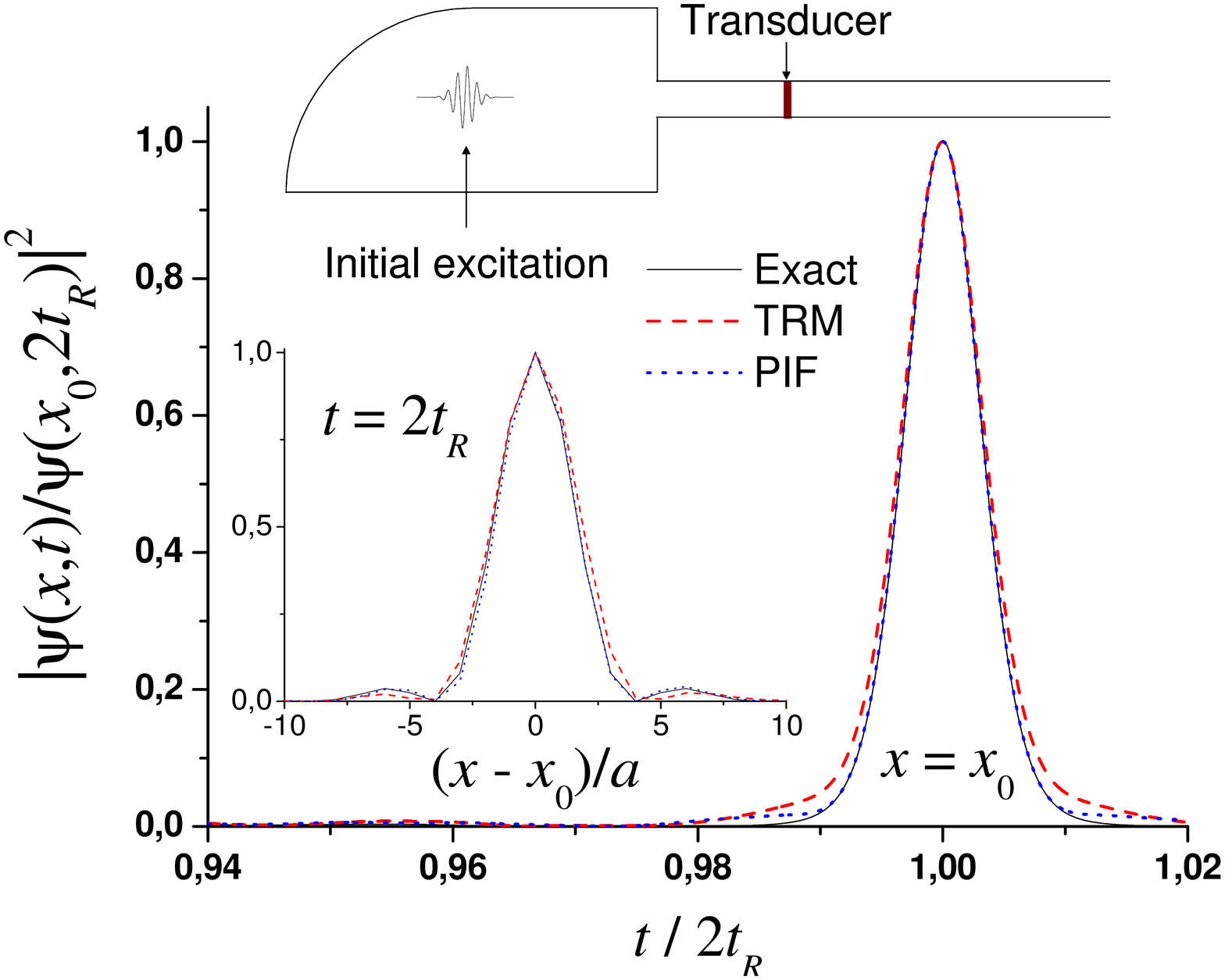}
\caption{(color online). Focalization at the focal point for TRM (red
dashed) and PIF (blue dotted) procedures contrasted with the exact reversal
(solid). PIF and exact curves coincide for times close to $2t_R$. Top: basic
scheme of a quantum bazooka device with a localized excitation inside. Left
inset: spatial contrast in the recovered signal.}
\label{fig_1}
\end{figure}

In order to evaluate the eq.(\ref{eq_PIF}) we propose a model for a quantum
bazooka device composed by a stadium billiard coupled with a one-dimensional
waveguide as shown in the top of fig.(\ref{fig_1}). The discretization in
the Schr\"{o}dinger equation gives the tight-binding structure of the
Hamiltonian%
\begin{equation}
\hat{H}=\hat{H}_{B}+\hat{H}_{BC}+\sum_{j}\left( E_{j}\hat{c}_{j}^{\dag }\hat{%
c}_{j}+V_{j,j+1}\hat{c}_{j+1}^{\dag }\hat{c}_{j}+\mathrm{h.c.}\right) ,
\end{equation}%
where $\hat{H}_{B}$ is the Hamiltonian of the billard in a discrete basis
and $\hat{H}_{BC}$ the coupling with the waveguide. In the waveguide, $E_{j}$
denotes the energy at site $x_{j}=ja$ and $V_{j,j+1}$ is the hopping
amplitude between sites $x_{j}$ and $x_{j+1}$. Supposing a single
semi-infinite waveguide, the cavity results delimited by a single transducer
at site $x_{s}$. The Green's function is obtained through the continued
fraction technique as%
\begin{equation}
G_{s,s}(\varepsilon )=\frac{1}{\varepsilon -E_{s}-\Sigma _{\mathrm{in}%
}(\varepsilon )-\Sigma _{\mathrm{out}}(\varepsilon )},
\end{equation}%
where $\Sigma _{\mathrm{in}}(\varepsilon )$ and $\Sigma _{\mathrm{out}%
}(\varepsilon )$ are the self-energy corrections due to the presence of
sites inside and outside the cavity respectively \cite{Medina}. Notice that
the presence of the billard is included through $\Sigma _{\mathrm{in}%
}(\varepsilon )$.

Decimation on the cavity gives the $\Sigma _{\mathrm{in}}(\varepsilon )$
contribution as a continued fraction composed by the hoppings and the site
energies. In absence of magnetic fields or dissipation those parameters are
all real numbers. Therefore,%
\begin{equation}
\Sigma _{\mathrm{in}}(\varepsilon )=\frac{V_{s,s-1}V_{s-1,s}}{\varepsilon
-E_{s-1}-\dfrac{V_{s-1,s-2}V_{s-2,s-1}}{\varepsilon -E_{s-2}-\cdots \tfrac{%
V_{1,0}V_{0,1}}{\varepsilon -E_{0}-...}}},
\end{equation}%
is also a real function regardless the details of $H_{B}$.

On the other hand, in the limit where the number of sites increases
indefinitely, the homogeneous outer region, a linear chain with site
energies $E_{\mathrm{o}}$ and hoppings $V$, contributes to the self energy
as a complex function \cite{Medina}:%
\begin{equation}
\Sigma _{\mathrm{out}}(\varepsilon )=\Delta (\varepsilon )-\mathrm{i}\Gamma
(\varepsilon ).
\end{equation}%
The relationship between $\Gamma (\varepsilon )$ and the group velocity can
be found through the dispersion relation for the asymptotic waves:%
\begin{align}
\varepsilon _{k}& =E_{\mathrm{o}}-2V\cos (ka), \\
v_{g}& =\frac{1}{\hbar }\frac{\mathrm{d}\varepsilon _{k}}{\mathrm{d}k}=\frac{%
2Va}{\hbar }\sin (ka)=v_{\max }\sin (ka) \\
& =\frac{2a}{\hbar }\sqrt{V^{2}-\left( \frac{\varepsilon _{k}-E_{\mathrm{o}}%
}{2}\right) ^{2}}=\frac{2a}{\hbar }\Gamma (\varepsilon _{k}).
\end{align}%
We choose $E_{\mathrm{o}}=2V$ so that the differential form of the Schr\"{o}%
dinger equation for a particle with mass $m$ is obtained as the limit $\hbar
^{2}/2Va^{2}\longrightarrow m$ when $V\rightarrow \infty $ and $%
a^{2}\rightarrow 0$. Since only the imaginary component from $\Sigma _{%
\mathrm{out}}(\varepsilon )$ is required, eq.(\ref{eq_PIF}) writes%
\begin{equation}
\chi _{s}^{\mathrm{PIF}}(\varepsilon )=\frac{2}{\hbar }e^{\mathrm{i}%
\varepsilon 2t_{R}/\hbar }\Gamma (\varepsilon )\psi _{s}^{\ast }(\varepsilon
).
\end{equation}%
The complex exponential serves to define the origin of time, and hence we
can neglect it. Using the definition of $\Gamma (\varepsilon )$ given above
we get%
\begin{align}
\chi _{s}^{\mathrm{PIF}}(\varepsilon )& =\sqrt{1-\left( \frac{\varepsilon -2V%
}{2V}\right) ^{2}}\tfrac{2V}{\hbar }\psi _{s}^{\ast }(\varepsilon ) \\
& =\frac{v_{g}(\varepsilon )}{v_{\max }}\chi _{s}^{\mathrm{TRM}}(\varepsilon
).  \label{eq_final}
\end{align}%
The exact reversal requires the injection of a TRM signal filtered by the
group velocity of the scattered waves. When the initial state is a local
excitation composed by a few sites inside the cavity, the Fourier transform
of the detected signal $\psi _{s}(\varepsilon )$ will cover the whole energy
band. The factor in the PIF procedure shows that in dispersive media the
correction is effective near the band edges where the group velocity becomes
negligible.

In the numerical solution, the propagation of an initial gaussian
wave-packet with enegy $(k_{o}a)^{2}V$ is recorded by a single transducer
placed in the waveguide consisting of a single propagating mode (in
practice, this waveguide is well represented by a one-dimensional chain).
The time evolution was performed through a second order Trotter-Suzuki
algorithm \cite{DeRaedt}. Fig.(\ref{fig_1}) shows the recovered signal at
the focal point, \textit{i.e.} the centroid of the original wave-packet. The
focalization functions for both methods are close to the exact reversed
propagation. However, a slight broadening of the focalized wave-packet is
observed in the TRM. In this case, the differences between PIF and TRM
procedures are small because the initial excitation is mainly composed by
states whose group velocity in the outgoing channel remains almost constant.

\subsection{PIF in a classical Helmholtz resonator}

The PIF procedure for a classical wave equation can be readily deduced using
our Green's function strategy by considering its finite difference version.
This is best visualized as a system of coupled oscillators \cite{Economou}.
By including an inhomogeneity, it becomes in a simple model for a Helmholtz
resonator coupled to an acoustic waveguide. Here, the lowest frequency $%
\omega _{0}$ of the $\lambda /2$ mode in the resonator is represented by a
single mass $m_{0}$ and its corresponding spring.\ The waveguide is modelled
by a semi-infinite chain of identical masses $m=\alpha m_{0}$ placed at the
equilibrium points $x_{n}=na$ (with $a$ the lattice constant and $n$\ a
positive integer). The nearest neighbor spring constant is $K=m\omega _{%
\mathrm{x}}^{2}$. The equations of motion for the corresponding
displacements\ $u_{n}$ can be written in a matrix form which, in the
frequency domain, reads:%
\begin{equation}
\mathbb{D}^{-1}(\omega )\mathbf{u}(\omega )=\left( 
\begin{array}{ccc}
\omega ^{2}-\tilde{\omega}_{0}^{2} & \alpha \omega _{\mathrm{x}}^{2} & \cdots
\\ 
\omega _{\mathrm{x}}^{2} & \omega ^{2}-2\omega _{\mathrm{x}}^{2} &  \\ 
\vdots &  & \ddots%
\end{array}%
\right) \mathbf{u}(\omega )=0,
\end{equation}%
where $\tilde{\omega}_{0}^{2}=\omega _{0}^{2}+\alpha \omega _{\mathrm{x}%
}^{2} $. Here, $\mathbb{D(\omega )}$ represents a momentum-displacement
response function, which is the resolvent of the dynamical matrix.\ Any
diagonal element has a simple expression in terms of continued fractions. In
particular, a far-field transducer is placed at the site $x_{s}$ on the
waveguide, where the response function is%
\begin{equation}
D_{s,s}(\omega )=\frac{1}{\omega ^{2}-2\omega _{\mathrm{x}}^{2}-\Delta
_{L}(\omega )-\left[ \Delta _{R}(\omega )-\mathrm{i}\omega \eta (\omega )%
\right] }.
\end{equation}%
Here, the mean-field frequency $2\omega _{\mathrm{x}}^{2}$ appears shifted
by the dynamical effect from the oscillators at both sides of $x_{s}$. The
imaginary shift indicates that excitation components of different
frequencies would eventually escape through the waveguide at the right with
group velocities%
\begin{equation}
v_{g}(\omega )=\frac{a}{2}\eta (\omega )=\frac{a}{2}\sqrt{4\omega _{\mathrm{x%
}}^{2}-\omega ^{2}}.
\end{equation}%
The propagation of an excitation originated in the resonator $x_{0}$ is
detected in $x_{s}$ inside the waveguide. The recorded signal presents a
strong component in $\tilde{\omega}_{0}$, which is the \textquotedblleft
carrier\textquotedblright\ frequency as can be appreciated from the density
of states at the resonator's site. The displacement at the transducer,
resulting from an initial excitation $\xi _{0}$ in the resonator, can be
expressed as%
\begin{equation}
u_{s}(t)=G_{s,0}(t)\xi _{0},\ t>0,  \label{eq_reg}
\end{equation}%
where $G_{s,0}(t)$ is the Green's function describing the
displacement-displacement response. In general, the connection between the
Green's function and the momentum-displacement response is given by%
\begin{equation}
G_{i,j}(\omega )=\mathrm{i}\omega D_{i,j}(\omega ).
\end{equation}%
As before, the complete evolution $\tilde{u}_{s}(t)$ in the transducer can
be conceived as the forward evolution corresponding to the positive times
and a backward evolution accounting for the negative ones. In this sense,%
\begin{equation}
\tilde{u}_{s}(t)=\left\{ 
\begin{array}{cc}
u_{s}(-t), & t<0, \\ 
u_{s}(t), & t\geq 0,%
\end{array}%
\right.
\end{equation}%
and the injection that produces the desired reversion can be obtained as%
\begin{equation}
\delta u_{s}(\omega )=\frac{\tilde{u}_{s}^{\ast }(\omega )}{G_{s,s}(\omega )}%
.
\end{equation}

According to eq.(\ref{eq_reg}), the complete evolution writes in the
frequency domain as%
\begin{align}
\tilde{u}_{s}^{\ast }(\omega )& =u_{s}^{\ast }(\omega )+u_{s}(\omega ) \\
& =\left[ G_{s,0}^{\ast }(\omega )+G_{s,0}(\omega )\right] \Delta u_{0}.
\end{align}%
Here again, the key tool is the Dyson equation connecting the two subspaces
delimited by the transducer. We can write 
\begin{equation}
G_{s,0}(\omega )=\frac{G_{s,s}(\omega )}{G_{s,s}^{\ast }(\omega )}%
G_{s,0}^{\ast }(\omega ),
\end{equation}%
and the PIF injection rewrites in terms of the partial evolution
corresponding to the detected signal%
\begin{align}
\delta u_{s}^{\mathrm{PIF}}(\omega )& =\left[ \frac{1}{G_{s,s}(\omega )}+%
\frac{1}{G_{s,s}^{\ast }(\omega )}\right] u_{s}^{\ast }(\omega ) \\
& =\frac{1}{\mathrm{i}\omega }\left[ \frac{1}{D_{s,s}(\omega )}-\frac{1}{%
D_{s,s}^{\ast }(\omega )}\right] u_{s}^{\ast }(\omega ) \\
& =\eta (\omega )u_{s}^{\ast }(\omega ).
\end{align}

As in the quantum case, we denote $2\omega _{\mathrm{x}}u_{s}^{\ast }(\omega
)$ as the Fourier transform $\delta u_{s}^{\mathrm{TRM}}(\omega )$ of the
injection function in the TRM protocol. In consequence, the perfect time
reversal is obtained only once a further filter $v_{g}(\omega )/v_{\max }$
is applied. Hence, the PIF formula for internal source in the acoustic case
is%
\begin{equation}
\delta u_{s}^{\mathrm{PIF}}(\omega )=\frac{v_{g}(\omega )}{v_{\max }}\delta
u_{s}^{\mathrm{TRM}}(\omega ).
\end{equation}

Notably, the prescription remains exactly the same as that of the quantum
version (see eq.(\ref{eq_final})). This implies that the effectiveness of
the filter will depend on the structure of the waveguide and the initial
wave-packet, regardless the details of the cavity. However, for cases in
which the group velocity in the free space is constant, it is clear that $%
\delta u_{s}^{\mathrm{PIF}}(\omega )\equiv \delta u_{s}^{\mathrm{TRM}%
}(\omega )$.

\begin{figure}[tbp]
\includegraphics[angle=90,width=8cm]{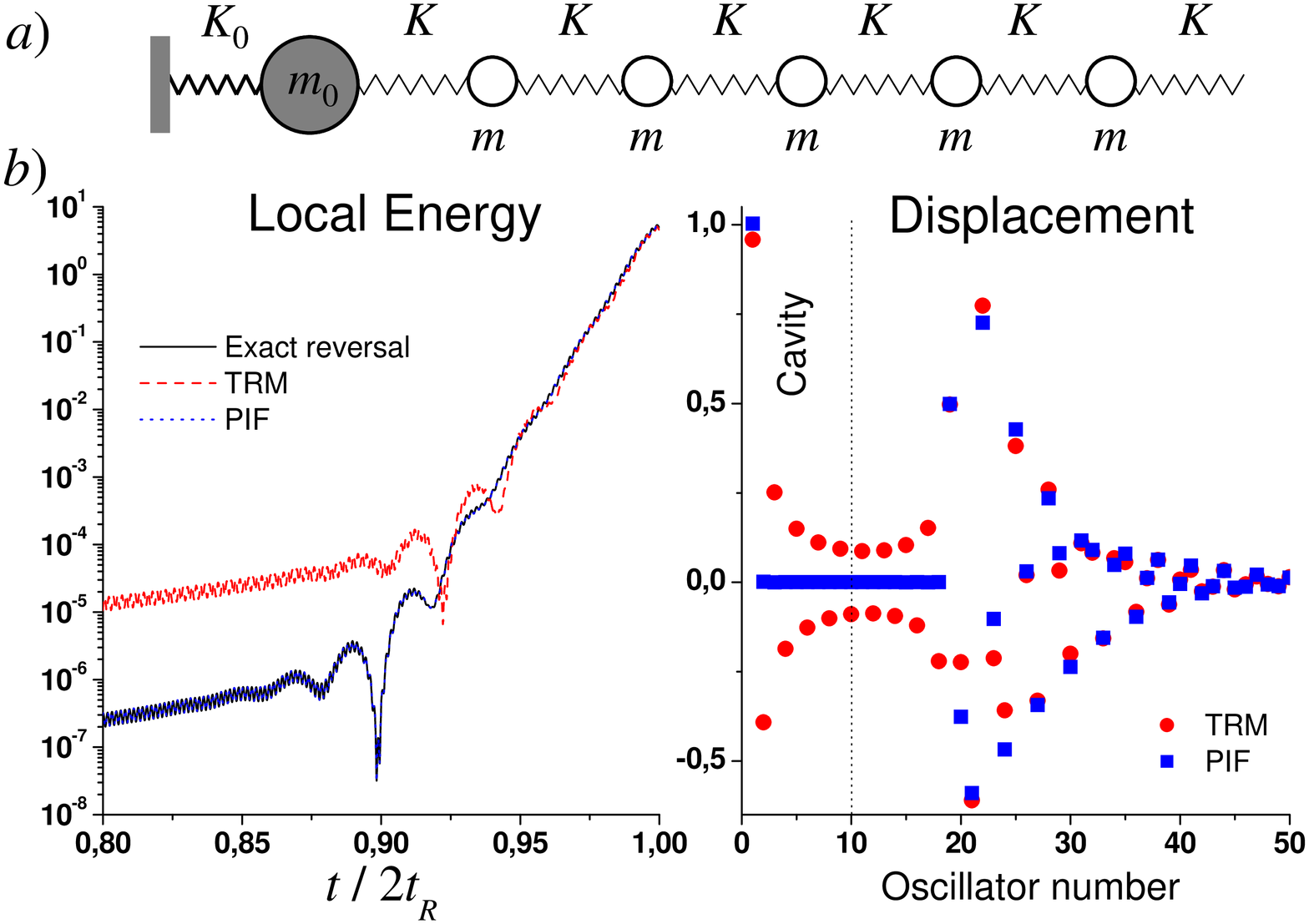}
\caption{(color online). (a) Balls and springs model for a Helmholtz
resonator coupled to a waveguide. (b) Focalization signal in the resonator.
Left panel: local energy recovered as function of time. Both methods, the
TRM (red dashed) and PIF (blue dotted) are contrasted with the exact
reversal (solid). PIF and exact curves coincide for the whole temporal
range. Right Panel: Displacements at the focalization time.}
\label{fig_2}
\end{figure}

A numerical simulation of the reconstruction of the initial excitation was
performed using the Pair Partitioning method (PPM) \cite{CalvoMC}, which
yields the complete dynamics by alternating among the evolutions of pairs of
coupled\ masses. In a similar way as the Trotter algorithm, PPM approximates
the actual evolution determined by the Hamiltonian $H_{1}+H_{2}$ during a
small time step $\delta t$ as a sequence of unitary transformations $U\left[
(H_{1}+H_{2})\delta t\right] \simeq U\left[ H_{1}\delta t\right] U\left[
H_{2}\delta t\right] $ and results in a perfectly reversible algorithm. The
calculated dynamics is best depicted, as in fig.(\ref{fig_2}), by analyzing
the local energy which avoids the fast fluctuations shown by displacement
and momentum. Here, the left panel shows (in a log scale) that the recovered
resonator's local energy coincides with the exact reversal over all the
relevant ranges. Since in this model the waveguide has a cut-off frequency,
as would be the case in a grated waveguide, the PIF filter improves the TRM
focusing when it comes to reproduce the low intensity signals. This is
particularly evident in the reproduction of the time-reversed \textit{%
survival collapse}. This is a sudden dip in the local energy resulting from
the interference between the excitation surviving in the resonator and that
returning from the waveguide. This surprising phenomenom was originally
described in the context of quantum spin channels \cite{Rufeil}. The perfect
contrast of the time reversed signal provided by PIF is evidenced in the
right panel by the exact cancellation of displacements except for the
resonator. It is also interesting to notice that both procedures produce a
phantom signal outside the \textquotedblleft silence
region\textquotedblright . While the TRM has an evident imperfection in the
localization of the signal, the PIF\ procedure yields this absolute
cancellation even outside the cavity region defined by the far field
transducer at $x_{s}$.\textbf{\ }The resulting localization, which
corresponds to $\lambda /2$ of the carrier signal, was only enabled by
filtering out the band edge components in the emitted signal. From this
perspective, the PIF procedure contributes to the goal of achieving focusing
beyond the diffraction limit \cite{FinkSCIENCE}.

\section{Conclusions}

We presented an expression of the PIF procedure which results particularly
simple when the excitation is generated by an internal source. We observed
that, contrary to what happens when the source is external, the prescription
does not involve internal details of the cavity but only the simpler
information about the propagation in the outer region. Hence, this filter
applies to two physically relevant situations: 1) When the excitation is
actually originated in the interior of the cavity. In this case, one obtains
a perfect reversal of the wave function for all the times after the source
has been turned-off. 2) When one uses an external source in a situation that
allows for a clear separation between incoming and outgoing waves. In that
case, one can use the recording of the outgoing wave to perfectly reverse
the whole excursion of the excitation through the cavity. This condition is
achieved when the boundaries are placed far enough from the reverberant
region, as in the quantum and acoustic bazooka devices. An interesting
consequence of our result in the acoustical case is that TRM produces a
perfect time reversal within the cavity, provided that the propagation
beyond its contour is free of further reverberances.

A numerical assesment of the reversion fidelity shows that PIF constitutes a
notable improvement over TRM prescription in cases where the energy
(frequency) dependence has a non-linear dispersion relation, as typical
excitations described by the Schr\"{o}dinger equation or, in acoustic
systems, when the escape velocity is not constant. This occurs in the
presence of multiple propagating channels or when collisions outside the
cavity have a relevant contribution as in grated waveguides.

The PIF prescription becomes an analytical tool to design specific
excitations well beyond the discussed TRM context, thus providing an
alternative strategy for coherent control. In particular, it allows the
design of quantum bazookas that shoot wave packet excitations. Feasible
implementations of this concept include the local generation of vibrational
waves in nanoelectromechanical structures \cite{Schwab}. There, the
excitation is injected through the coupling of a Cooper-pair box to a
nanocantilever. In Bose-Einstein condensates confined to an optical lattice 
\cite{Chu}, there is a well-defined set of quantum states with tunable
couplings. Thus, the generation of macroscopic wavefunctions would benefit
from our simple and consistent PIF prescription. Last but not least, in NMR,
the injection of targeted spin-wave packets in chains of interacting spins 
\cite{Madi-Ernst} is possible through the local injection of the
polarization stored in a rare $^{13}C$ nucleus \cite{Alvarez}.

\acknowledgments We thank R.A. Jalabert and L.E.F. Foa Torres for valuable
discussions, F. Vignon for helpful correspondence and F. Pastawski for
comments on the manuscript. We acknowledge financial support of CONICET,
ANPCyT and SECyT-UNC.

\end{document}